\documentclass[aps,prl,twocolumn,groupedaddress]{revtex4}

\usepackage{amsmath, amsthm,amssymb}
\usepackage{graphicx}
\usepackage{dcolumn}
\usepackage{bm}

\begin{document}


\title{Quench Dynamics of the Anisotropic Heisenberg Model}


\author{Wenshuo Liu}
\email[]{wenshuoliu@gmail.com}
\affiliation{Department of Physics and Astronomy, \\Rutgers University \\Piscataway, New Jersey 08854}
\author{Natan Andrei}
\email[]{natan@physics.rutgers.edu}
\affiliation{Department of Physics and Astronomy, \\Rutgers University \\Piscataway, New Jersey 08854}


\date{\today}

\begin{abstract}
We develop an analytic approach for the study of the quench dynamics of the anisotropic Heisenberg model (XXZ model) on the infinite line. We present the exact time-dependent wavefunctions after a quench in an integral form for any initial state and for any anisotropy $\Delta$ by means of a generalized Yudson contour representation. We calculate the evolution of several observables from two particular initial states: starting with a local N\`eel state we calculate the time evolution of the antiferromagnetic order parameter--staggered magnetization; starting with a state with consecutive flipped spins we calculate the propagation of magnons and bound state excitations, and the induced spin currents. We also show how the ``string'' solution of Bethe Ansatz equations emerge naturally from the contour approach. We confront our results with experiments and numerical methods where possible.
\end{abstract}

\pacs{}

\maketitle
The study of the time evolution of quantum many-body systems has seen much progress due to the advance of techniques in ultracold atomic gases, which can provide almost isolated systems with highly tunable parameters \cite{RevModPhys.80.885}. For example, the ``quantum Newton's cradle'' \cite{kinoshita2006quantum} lead to intense theoretical study of the relation between integrability and thermalization, or lack thereof \cite{PhysRevLett.98.050405, rigol2008thermalization, PhysRevLett.97.156403, PhysRevLett.98.210405, PhysRevLett.100.100601, fioretto2010quantum, PhysRevLett.98.180601, PhysRevLett.105.250401, PhysRevLett.106.227203, 1742-5468-2007-06-P06008}. A standard setup to study a system out of equilibrium is a quantum quench: one prepares the system in some initial state and then suddenly applies a Hamiltonian to it while monitoring its subsequent time evolution. In particular, the quench of a quantum spin chain drew much attention from different aspects \cite{PhysRevB.82.144302, PhysRevE.71.036102, 
1742-5468-2013-10-P10028, 1742-5468-2013-07-P07012}. The quench from the antiferromagnetic phase to the critical phase follows the evolution of the order parameter through a critical point \cite{barmettler2010quantum, fagotti2013dynamical}; the emergence of propagating bound states of magnons, which was predicted by Bethe Ansatz \cite{Bethe1931}, are studied by various numerical methods such as time-evolving block decimation \cite{PhysRevLett.108.077206}, and directly observed experimentally \cite{2013arXiv1305.6598F}. 

We shall study the quench dynamics of a quantum spin chain by an exact, analytical method: Yudson's contour approach \cite{yudson1988dynamics} proposed  in the context of the Dicke model. It is proved to be also successful in one-dimensional interacting boson gas \cite{PhysRevLett.109.115304, PhysRevA.87.053628} which has different dynamics but a similar scattering-matrix. Although the quantum spin chain system has a more complicated S-matrix, we will show the Yudson's contour approach works as well after some generalization. 

The physics of a quantum spin chain is given by the anisotropic Heisenberg model:
\begin{align}
 H=-J\sum_j\left[\sigma_j^x\sigma^x_{j+1}+\sigma^y_{j}\sigma^y_{j+1}+\Delta(\sigma^z_j\sigma^z_{j+1}-1)\right],\label{hamiltonian}
\end{align}
where the $\sigma$'s are Pauli matrices. We choose $J>0$ and measure time in dimensionless units $Jt$. The complete set of eigenstates is of the form \cite{Bethe1931, PhysRev.112.309, sutherland2004beautiful, takahashi2005thermodynamics}:
\begin{align}
 |\vec{k}\rangle=&\sum_{\{m_j\}}\mathcal{S}\prod_{i<j}[\theta(m_i-m_j)+s(k_i,k_j)\theta(m_j-m_i)]\notag\\
 &\times\prod_je^{ik_jm_j}\sigma_{m_j}^+|\Downarrow\rangle,\label{eigenk}
\end{align}
where $\sigma^+=(\sigma^x+i\sigma^y)/2$ is the spin raising operator, and $|\Downarrow\rangle$ is the reference state with all the spins down. $\mathcal{S}$ is the symmetrizer on the $m_j$'s. The interaction between magnons is  captured by the S-matrix: $s(k_i,k_j)=-\frac{1+e^{ik_i+ik_j}-2\Delta e^{ik_i}}{1+e^{ik_i+ik_j}-2\Delta e^{ik_j}}$. For  thermodynamic calculations it is convenient to impose periodic boundary conditions on the system so the momenta $k_j$'s are quantized as solutions of the resulting Bethe Ansatz equations. They may be either real or complex; the real ones are interpreted as free magnons, while the complex ones are realized as bound states \cite{sutherland2004beautiful, takahashi2005thermodynamics}. The bound states of magnons are difficult to detect by conventional probes, but were observed by means of quantum quench \cite{PhysRevLett.108.077206, 2013arXiv1305.6598F}. We will show how they emerge naturally from the Yudson's contour approach. 

\begin{table*}
 \caption{\label{parametrization}Different parameterizations for various $\Delta$. Instead of momentum $k$, the rapidity $\alpha$ is to label an eigenstate. All the physical quantities are expressed as a function of $\alpha$. The anisotropy $\Delta$ is denoted by a new parameter $\lambda$ or $\mu$.}
 \begin{ruledtabular}
 \begin{tabular}{c c c c c}
Physical quantities	&	In $k$ language	&	$\Delta=-\cosh\lambda<-1$	&	$-1<\Delta=-\cos\mu<1$	& 	$\Delta=\cosh\lambda>1$\\
 \hline
Plane wave $P^m(\alpha)$	&	$e^{ik m}$	&	$\left[\frac{\sin\frac{i\lambda-\alpha}{2}}{\sin\frac{i\lambda+\alpha}{2}}\right]^m$	&	$\left[\frac{\sinh(\frac{i\mu-\alpha}{2})}{\sinh(\frac{i\mu+\alpha}{2})}\right]^m$	&	$\left[-\frac{\sin\frac{i\lambda-\alpha}{2}}{\sin\frac{i\lambda+\alpha}{2}}\right]^m$\\

S-matrix $S(\alpha_1,\alpha_2)$	&	$-\frac{1-e^{ik_1+ik_2}-2\Delta e^{ik_1}}{1-e^{ik_1+ik_2}-2\Delta e^{ik_2}}$	&	$\frac{\sin(\frac{\alpha_1-\alpha_2}{2}-i\lambda)}{\sin(\frac{\alpha_1-\alpha_2}{2}+i\lambda)}$	&	$\frac{\sinh(\frac{\alpha_1-\alpha_2}{2}-i\mu)}{\sinh(\frac{\alpha_1-\alpha_2}{2}+i\mu)}$	&	$\frac{\sin(\frac{\alpha_1-\alpha_2}{2}-i\lambda)}{\sin(\frac{\alpha_1-\alpha_2}{2}+i\lambda)}$\\

Eigen energy $E(\alpha)$	&	$4J(\Delta-\cos k)$	&	$\frac{4J\sinh^2\lambda}{\cos\alpha-\cosh\lambda}$	&	$-\frac{4J\sin^2\mu}{\cosh\alpha-\cos\mu}$	&	$-\frac{4J\sinh^2\lambda}{\cos\alpha-\cosh\lambda}$\\

Weight $W(\alpha)=\frac{dk}{d\alpha}$	&	&	$\frac{\sinh\lambda}{\cosh\lambda-\cos\alpha}$	&	$\frac{\sin\mu}{\cosh\alpha-\cos\mu}$	&	$\frac{\sinh\lambda}{\cosh\lambda-\cos\alpha}$\\
 \end{tabular}
 \end{ruledtabular}
\end{table*}

\begin{figure*}
 \centering
\includegraphics[width=0.85\textwidth]{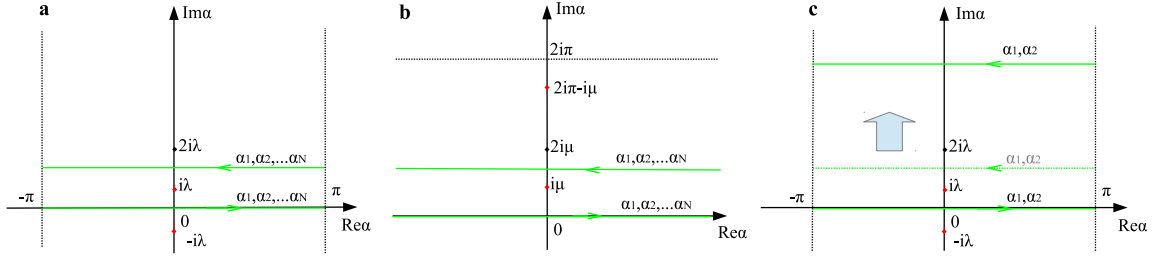}
\caption{The integral contours  for the regions: (a) $|\Delta|>1$, the integrand is periodic in the real direction with period $2\pi$. It has a pole at $\alpha_j^*=i\lambda$ or $-i\lambda$ from the plane wave depending on the sign of $m_j-n_j$; and a possible pole at $\alpha_j^*=\alpha_k+2i\lambda,k<j$ or $\alpha_k-2i\lambda,k>j$ from the S-matrices. (b) $|\Delta|<1$, the integrand is periodic in the imaginary direction with period $2i\pi$. The integrand has similar pole structure. (c) By moving the back-running part of the contours over the S-matrix pole at $\alpha_2^*=\alpha_1+2i\lambda$, the bound state emerges from the integral Eq. \eqref{wavefunction}.}
\label{contours}
\end{figure*}

To carry out the time evolution of some generic initial state $|\Psi_0\rangle$ one expands it by the Hamiltonian's eigenstates and evolves each one separately:
\begin{align}
&|\Psi_0\rangle=\sum_{\{k_j\}}|\vec{k}\rangle\langle \vec{k}|\Psi_0\rangle,\\
&|\Psi(t)\rangle=e^{-i\hat{H}t}|\Psi_0\rangle=\sum_{\{k_j\}} e^{-i\sum_jE(k_j)t}|\vec{k}\rangle\langle \vec{k}|\Psi_0\rangle.
\end{align}
Due to the complexity of the eigenstate $|\vec{k}\rangle$, both the overlap $\langle \vec{k}|\Psi_0\rangle$ and the summation over $k_j$'s are difficult to calculate. Yudson's contour approach allows us to carry out these calculations efficiently. One considers basis states defined in one quadrant $|n_1,\cdots, n_N\rangle = \theta(n_1>\cdots>n_N) \prod_{j=1}^N \sigma^+_{n_j}|\Downarrow\rangle$ in terms of which any initial state $|\Psi_0\rangle$ can be expressed and works directly on the infinite line, replacing the summation over $k_j$'s by integrals in the complex plane:
\begin{align}
|n_1,\cdots, n_N\rangle=\int_{\gamma_j}d\vec{k}|\vec{k}\rangle( \vec{k}|n_1,\cdots, n_N\rangle.\label{expanding}
\end{align}
Here $|\vec{k})=\sum_{\{m_j\}}\theta(m_1>\cdots>m_N)\prod_j e^{ik_j m_j}\sigma_{m_j}^+|\Downarrow\rangle$ is the Yudson's auxiliary state \cite{yudson1988dynamics}, whose overlap with an initial state is simple. The integral contours $\{\gamma_j \}$ need to be properly chosen, so that the identity \eqref{expanding} is mathematically valid. 
To exhibit the contours it is convenient to reparameterize the momenta $k$, the reparameterization being dependent on the value of the anisotropy $\Delta$ as given by Table I, in terms of which the contours $\{\gamma_j\}$ are shown in Fig. \ref{contours}(a)(b). The validity of the representation can be established by closing the contours in a carefully chosen order so that the integrand either has no pole inside the contour, or has poles inside the contour but the residues of those poles cancel out. The choice of contours  captures the properties of the bound states of the magnons, as we will show later. With the expansion \eqref{expanding}, we obtain the  time-dependent state,
\begin{align}
 |n_1,\cdots,n_N, t\rangle&=\int_{\gamma}d\vec{k}e^{-i\sum_jE(k_j)t}|\vec{k}\rangle(\vec{k} |n_1,\cdots,n_N \rangle\notag\\
& =\sum_{\{m_j\}}\Psi^{\{n_j\}}(\{m_j\},t)\prod_j\sigma_{m_j}^+|\Downarrow\rangle,
\end{align}
where the time-dependent wavefunction $\Psi^{\{n_j\}}(\{m_j\},t)$ is given in integral form,
\begin{align}
 \Psi^{\{n_j\}}&(\{m_j\},t)= \mathcal{S} \int_{\gamma}\prod_jd\alpha_jW(\alpha_j)P^{m_j-n_j}(\alpha_j) e^{-iE(\alpha_j)t} \notag\\
 \times &\prod_{i<j}[\theta(m_i-m_j)+S(\alpha_i,\alpha_j)\theta(m_j-m_i)].\label{wavefunction}
\end{align}
 The expression is valid for both regions $|\Delta|<1$ and $|\Delta|>1$, with different definitions of the functions $W(\alpha)$, $P^m(\alpha)$, $E(\alpha)$ and $S(\alpha_1, \alpha_2)$ given in Table \ref{parametrization}.

The time evolution \eqref{wavefunction} reveals the spectrum and eigenstate structure of the Hamiltonian which consists of propagating magnons and their bound states as well as bound states of bound states. Consider an example of an initial  state with two  spins flipped at $n_1,n_2$, propagating  with the $|\Delta|>1$ Hamiltonian. As shown in Fig. \ref{contours}(c), moving the back-running parts of the contours up to $+\infty i$ where they vanish, one picks up a pole contribution from the S-matrix at $\alpha_2^*=\alpha_1+2i\lambda$. Hence, the wavefunction can be separated into two terms
\begin{align}
 \Psi(m_1,m_2,t)
=\int_{-\pi}^{\pi}\prod_{j=1,2}d\alpha_jW(\alpha_j)P^{m_j-n_j}(\alpha_j)e^{-iE(\alpha_j)t}\notag\\
 \times[\theta_{m_1,m_2}+S(\alpha_1,\alpha_2)\theta_{m_2,m_1}]\notag\\
 +4\pi\sinh(2\lambda)\theta_{m_2,m_1}\int_{-\pi}^{\pi}d\alpha W(\alpha-i\lambda)P^{m_1-n_1}(\alpha-i\lambda)\notag\\
 \times e^{-iE(\alpha-i\lambda)t}W(\alpha+i\lambda)P_{m_2-n_2}(\alpha+i\lambda)e^{-iE(\alpha+i\lambda)t}.\label{boundstates}
\end{align}
The first term describes the propagation of free magnon contribution, while the second term, consisting of complex two-strings, describes the the propagation of the bound states contribution to the initial state of two flipped spins. This separation works for any number of flipped spins and in both regions $|\Delta|<1$ and $|\Delta|>1$. With more flipped spins, one needs to carefully count all the residues, and all the possible higher order strings will emerge \footnote{In the region $|\Delta|<1$, one moves the back-running contours to $i\pi$ instead of $+i\infty$, so the order of strings generated is restricted by the value of the anisotropy $\mu$. This feature agrees with the traditional thermodynamic Bethe Ansatz.}. Our method shows that the bound states are independent of boundary conditions, but intrinsic for the quantum dynamics of the spin chain. It also shows that the free magnon state and different bound states evolve separately. The prefactor in \eqref{boundstates}, given by the residue of the S-
matrix, can be interpreted as the proportion of bound eigenstate in the initial state. From the wavefunction any observable can be calculated, $\langle \hat{O}(t)\rangle =\langle\Psi(t)|\hat{O}|\Psi(t)\rangle$. For example, the local magnetization is given by $M(n,t)=\langle\Psi(t)|\sigma_n^z|\Psi(t)\rangle=\sum_{\{m_j\},j\neq 1}|\Psi(n,m_2,m_3,...,m_N,t)|^2$. Below we show our results of time-dependent observables including local magnetization, antiferromagnetic order parameter, and spin currents.  

\begin{figure*}
 \centering
\includegraphics[width=0.85\textwidth]{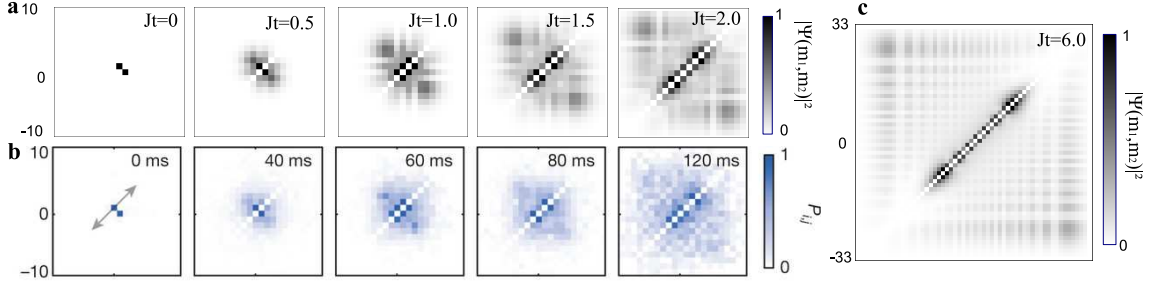}
\caption{(a) The norm of the wavefunction $|\Psi(m_1,m_2,t)|^2$ at different time for two flipped spins initially at $n_1=1$, $n_2=0$, calculated by numerical integration. (b) The joint probabilities of having two spins at site $i$ and $j$, measured experimentally in Ref. \cite{2013arXiv1305.6598F}(with permission). (c) At large time $Jt=6$, $|\Psi(m_1,m_2,t)|^2$ shows the features given by saddle point approximation.}
\label{twospins}
\end{figure*}

\begin{figure}
 \centering
\includegraphics[width=0.45\textwidth]{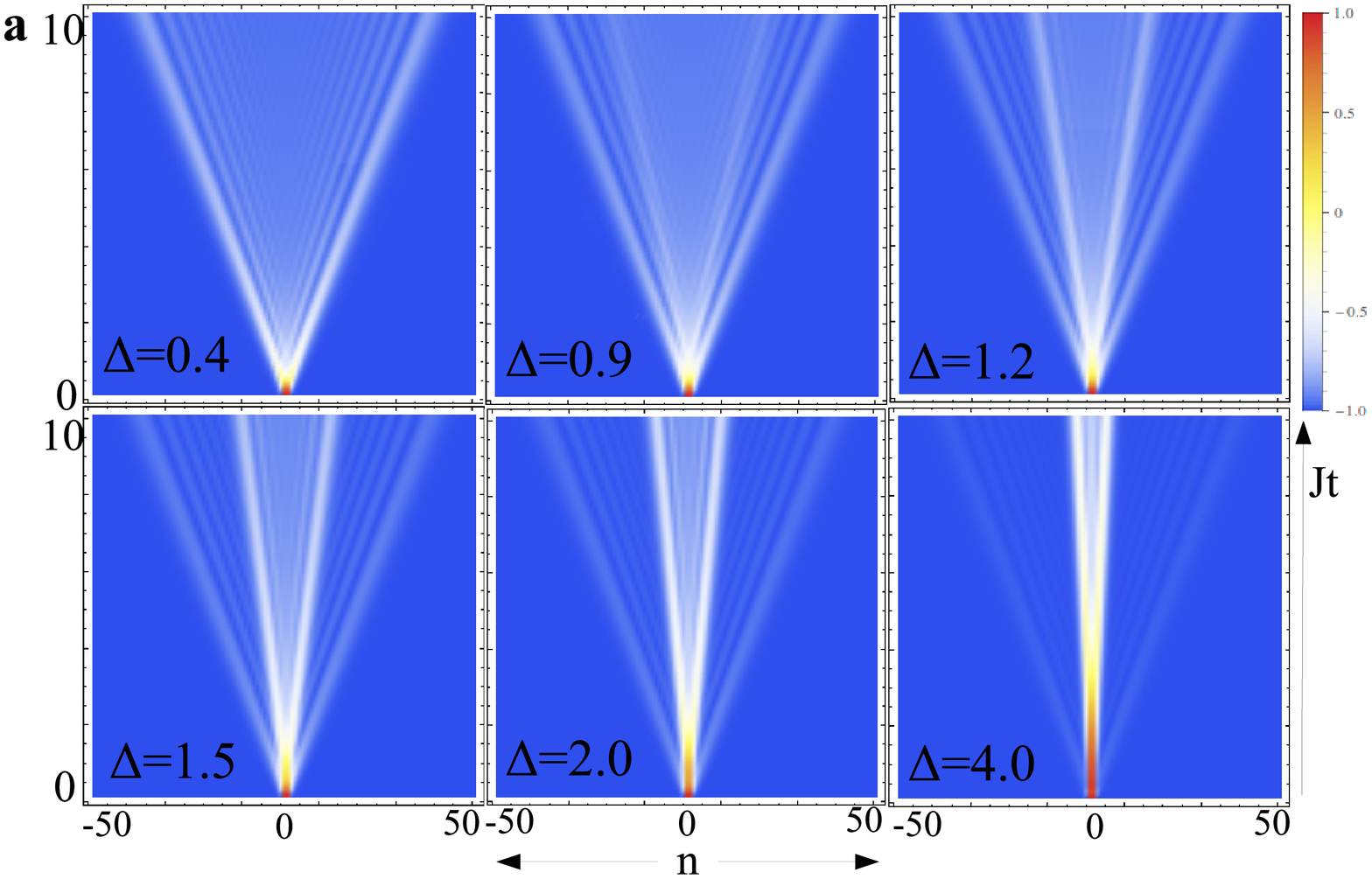}
\includegraphics[width=0.45\textwidth]{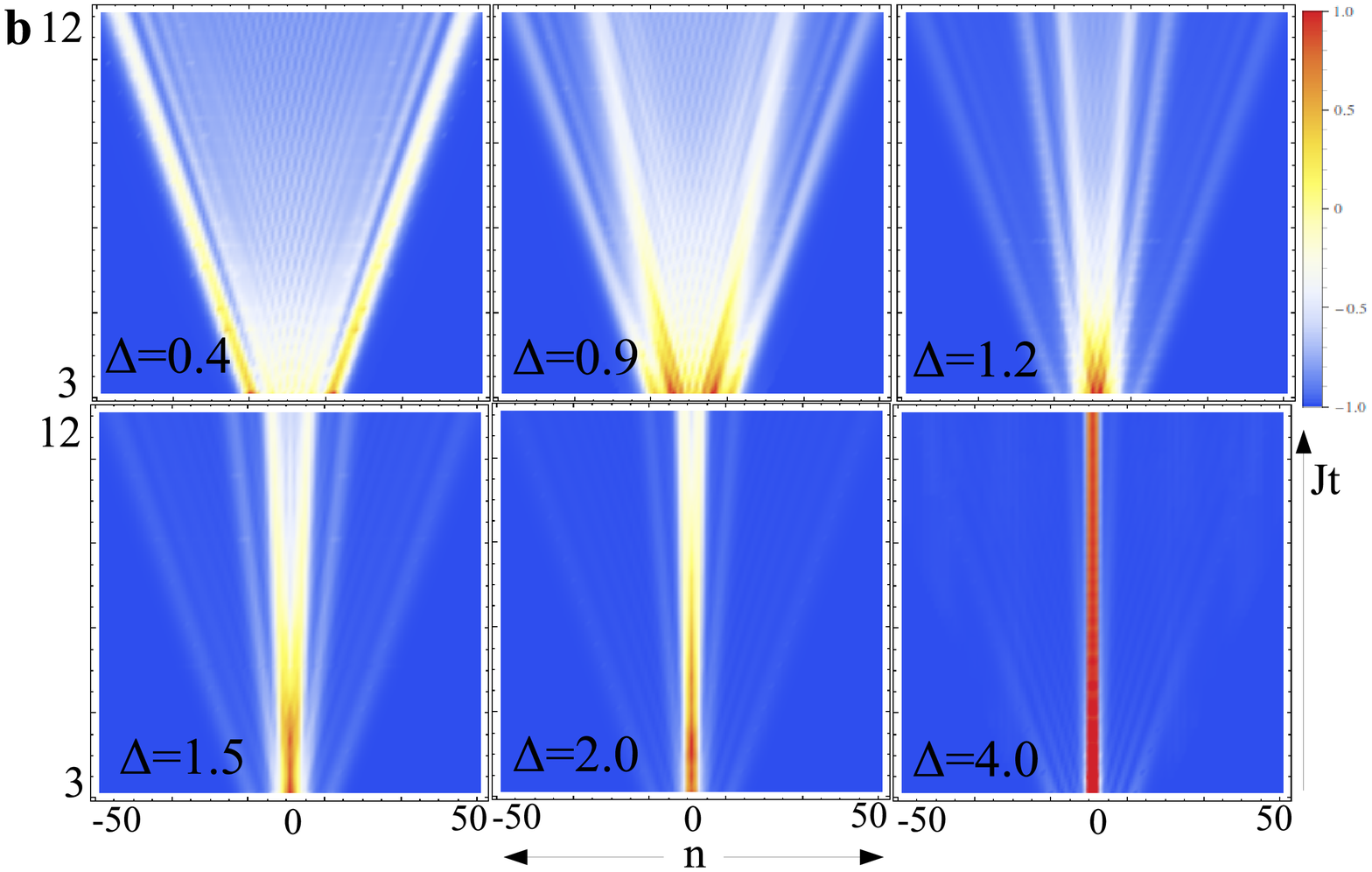}
\caption{Local magnetization as a function of site position $n$ and time $t$ for different $\Delta$, starting from an initial state (a) $|\Psi_0\rangle=\sigma_1^+\sigma_0^+|\Downarrow\rangle$, (b) $|\Psi_0\rangle=\sigma_1^+\sigma_0^+\sigma_{-1}^+|\Downarrow\rangle$.}
\label{localmag3}
\end{figure}

Generally, the integrals in \eqref{wavefunction} are difficult to calculate analytically but can be easily evaluated numerically for short times. For long times, the oscillatory integrand makes the numerical integration time-consuming, but allows on the other hand the evaluation of the integrals via the saddle point approximation. Consider the evolution of the initial state $|\Psi_0\rangle=\sigma_1^+\sigma_0^+|\Downarrow\rangle$ which was investigated experimentally \cite{2013arXiv1305.6598F}. Separating the contributions of propagating magnons and  bound states, the magnon contribution in $k$ language takes the form $\Psi_{magn}=\int_{-\pi}^{\pi}\prod_{j=1,2}\left[\frac{dk_j}{2\pi}e^{4iJt\cos k_j+i(m_j-n_j)k_j}\right]S(k_1,k_2)$ \footnote{For the region $m_1>m_2$, there is actually another term $\Psi_{magn}=\int_{-\pi}^{\pi}\prod_{j=1,2}\left[\frac{dk_j}{2\pi}e^{4iJt\cos k_j+i(m_j-n_j)k_j}\right]$. But it is trivial, since the the integral without an S-matrix can be decoupled for each $k_j$.}. For $Jt$ is 
large, it can be approximated by the contributions at the saddle points $\Psi_{magn}\cong\sum_{\{k_j^s\}}\psi(m_1,t)\psi(m_2,t)S(k_1^s,k_2^s)$, where the saddle points are determined by $\frac{\partial}{\partial k_j^s}(4iJt\cos k_j^s+i(m_j-n_j)k_j^s)=0\label{saddleeq}$, and $\psi(m_j,t)=\int_{-\pi}^{\pi}\frac{dk_j}{2\pi}e^{4iJt\cos k_j+i(m_j-n_j)k_j}=i^{m_j-n_j}J^B_{m_j-n_j}(4Jt)$ gives the propagation of a single magnon. In addition, when $m_1\approx m_2$ we have $k_1^s\approx k_2^s$ and since  $S(k,k)=-1$ we have $\Psi_{magn}(m_1\approx m_2)\approx 0$ along the diagonal. To describe  the bound state contribution we introduce the diagonal distance $M=(m_1-n_1+m_2-n_2)/2$, and the off-diagonal distance $N=(m_2-n_2-m_1+n_1)/2$. Then the bound state in $k$ language can be written as 
\begin{align}
 \Psi_{bound}^{n_1,n_2}(m_1,m_2,t)=&\theta_{m_2,m_1}\int_{-\pi}^{\pi}\frac{dK}{2\pi}e^{4iJt\frac{\sinh(\lambda)}{\sinh(2\lambda)}\cos K+iMK}\notag
 \\\times&(\cos K+\cosh 2\lambda)g(K)^N\label{boundstatek},
\end{align}
where $g(K)=\frac{1-\cos K}{2\cosh^2\lambda}$ takes real values between 0 and 1. The bound state looks like a single propagating magnon along the diagonal with a rescaled time, multiplying a exponential decay factor along the off-diagonal. In Fig. \ref{twospins} we show,  $P_{m_1,m_2}=|\Psi(m_1,m_2,t)|^2$,  the joint probability to have two spins at sites $m_1$ and $m_2$ for $\Delta=1.2$. The same quantity was experimentally measured \cite{2013arXiv1305.6598F}, reproduced in Fig. \ref{twospins}(b). The probability at long times, Fig. \ref{twospins}(c), exhibits the features discussed above. In Fig. \ref{localmag3}(a) we show the local magnetization, $M(n,t)$, as a function of site position $n$ and time $t$ for different $\Delta$. In the region $\Delta>1$ one can clearly see two wavefronts. The inner one is due to the bound state and the outer one is of free magnons. The overlap between the bound state and the initial state increases with $\Delta$ and for $\Delta=4$ the bound state dominates. The ratio of the 
velocities follows from the asymptotic form of the bound state \eqref{boundstatek} and is given by $\sinh(\lambda)/\sinh(2\lambda)$ since time is rescaled by this factor \footnote{According to the the dispersion relation of a m-string\cite{takahashi2005thermodynamics}, the maximum velocities are given by $v=\frac{dE(K)}{dK}|_{K=\frac{\pi}{2}}=4J\frac{\sinh(\lambda)}{\sinh(m\lambda)}, |\Delta|>1\label{boundvelocity}$}. For $\Delta<1$, it is harder to distinguish the second wavefront, because the bound state in this region takes a similar form as \eqref{boundstatek} with momentum $K$  restricted by $-\pi+2\mu<K<\pi-2\mu$.  Hence one of the saddle points could be out of the this range, making the wavefront of the bound state more diffuse. Especially when $\mu<\pi/4$ ($\Delta<\sqrt{2}/2$), even the momentum corresponding to the maximum velocity, $K=\pi/2$, is forbidden, so the wavefront completely disappears. 

Proceeding to  initial states with three flipped spins we consider first the initial state with consecutively flipped spins as the bound state contribution is maximal. Separating the wave function into magnon  and bound state contributions we calculate them by the saddle point approximation for $Jt>3$. The propagations are shown in Fig. \ref{localmag3}(b). For $\Delta>1$, one can clearly see three wavefronts: the outermost corresponds to the free magnons; the middle one is corresponding to the state with two magons bound while the third one is free; the innermost one corresponds to the state with three spins bounded together. Again, as $\Delta$ decrease below 1, the bound state wavefronts dim and become indistinct. 
\begin{figure}
 \centering
 \begin{tabular}{c}
\includegraphics[width=0.45\textwidth]{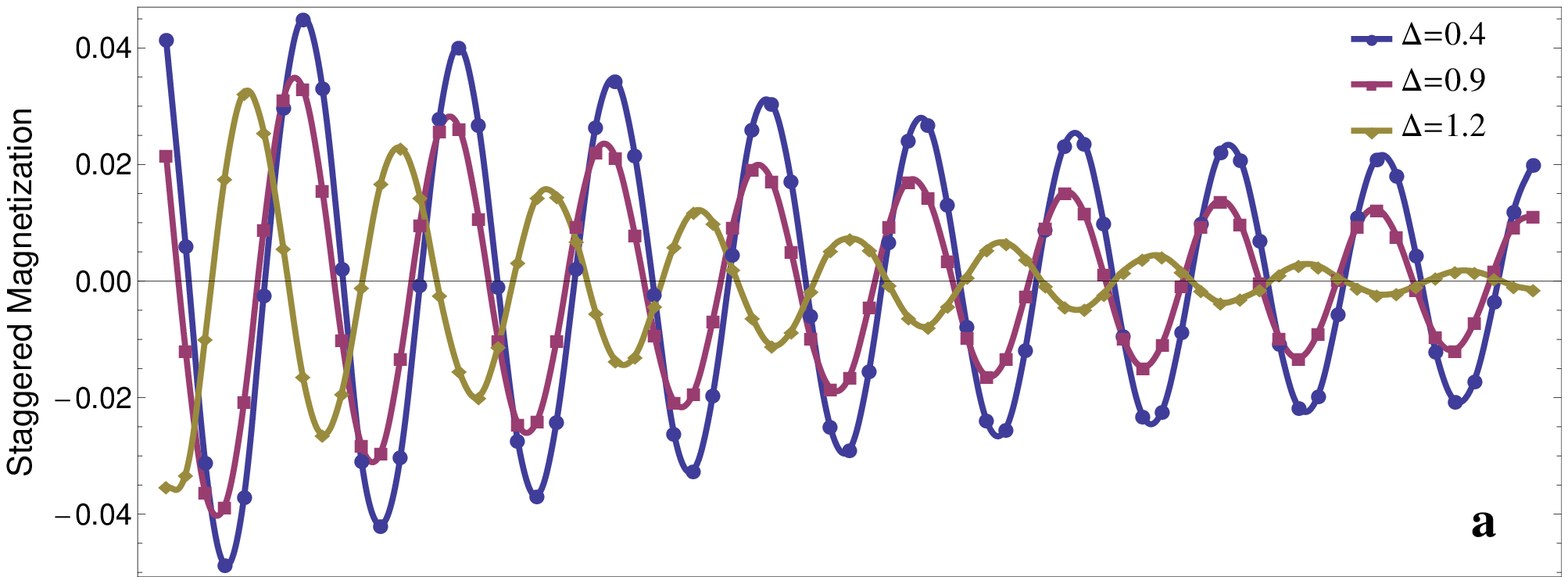}\\
\includegraphics[width=0.45\textwidth]{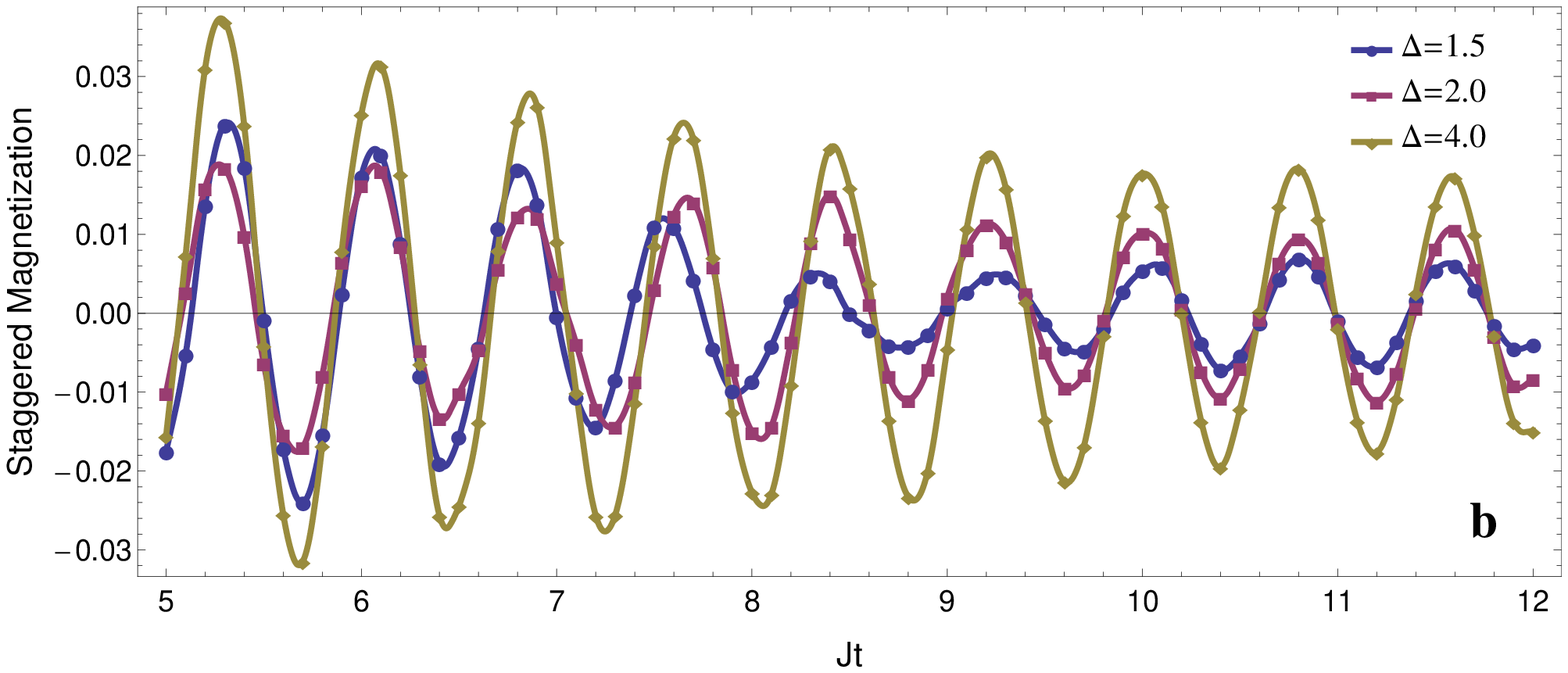}\\
 \end{tabular}
\caption{The time evolution of the staggered magnetization from an initial state $|\Psi_0\rangle=\sigma_2^+\sigma_0^+\sigma_{-2}^+|\Downarrow\rangle$, and within a box from site $n=-3$ to $n=3$. The curves connecting data points are quadratic spline interpolation.}
\label{mst}
\end{figure}

\begin{figure}
 \centering
 \begin{tabular}{c}
\includegraphics[width=0.45\textwidth]{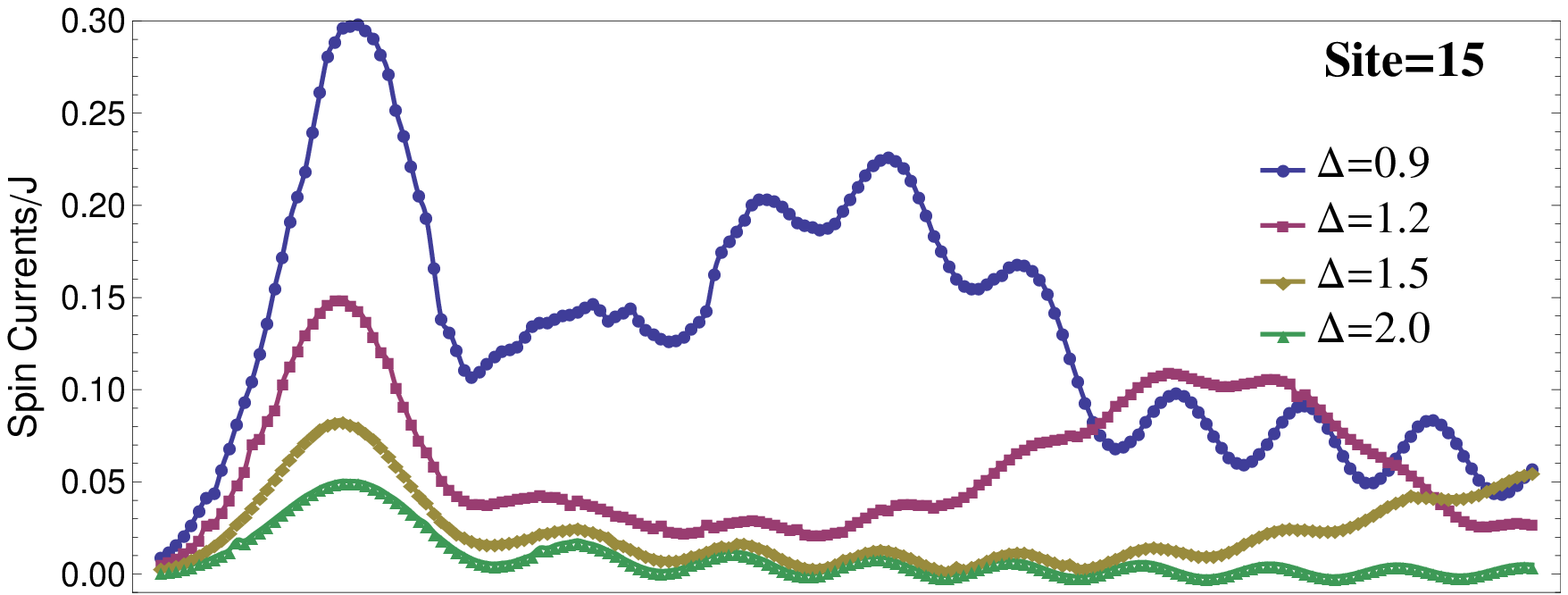}\\
\includegraphics[width=0.45\textwidth]{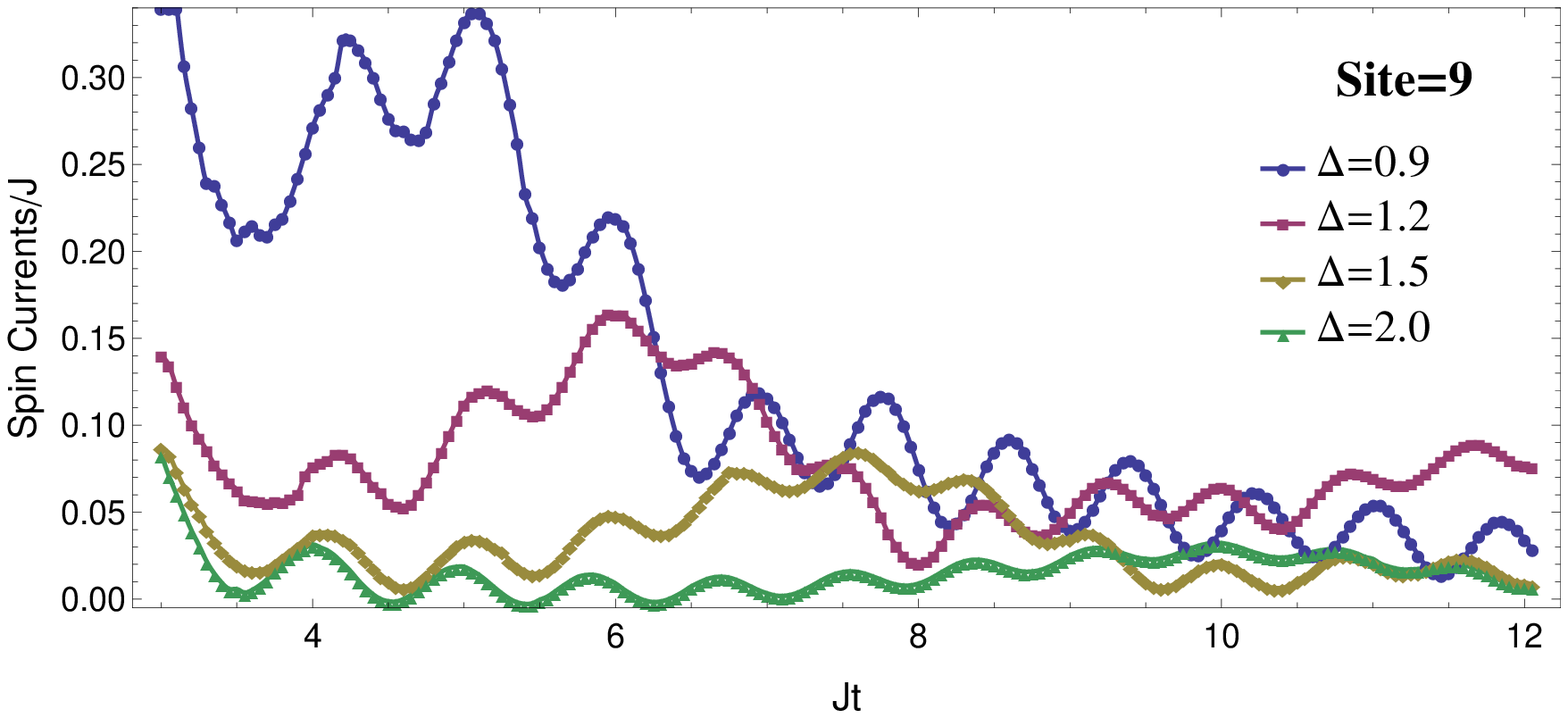}
 \end{tabular}
\caption{The spin currents measured at (a) site $n=15$ and (b) site $n=9$, from the initial state $|\Psi_0\rangle=\sigma_1^+\sigma_0^+\sigma_{-1}^+|\Downarrow\rangle$. One can see bumps caused by different bound states.}
\label{current}
\end{figure}
Of particular interest is the quench from an initial state exhibiting antiferromagnetic order (essentially corresponding to $\Delta= \infty$) to the critical phase $|\Delta|<1$ or finite $|\Delta|>1$. Starting from a local N\'eel state: $|\Psi_0\rangle=\prod_{j=1}^N\sigma_{2j}^+|\Downarrow\rangle$ for $N=3$ we follow the evolution of the staggered magnetization $m_s(t)=\sum_n(-1)^n\langle\sigma^z_n(t)\rangle\label{mags}$, the antiferromagnetic order parameter. The results calculated by our contour approach are shown in Fig. \ref{mst} for different $\Delta$. One can see typically both oscillation and decay of $m_s(t)$. In the critical regime, increasing  $\Delta$ leads to a faster decay of $m_s(t)$, while for $\Delta>1$, $m_s(t)$ decays more slowly as $\Delta$ increases. These decay rates roughly agree with Ref. \cite{barmettler2010quantum}, but the oscillation behavior does not. It may be due to the fact that we evaluate the order parameter in a finite box, and do not consider the magnons that get away. 

The quench induces spin currents in the system. The current through a site $n$, given by $\hat{I}_n=2iJ(\sigma_n^+\sigma_{n+1}^--\sigma_n^-\sigma_{n+1}^+)$, when evaluated as a function of time provides a measure of the various excitations that cross the site as they arrive at different times having different velocities. We show the currents from initial state $|\Psi_0\rangle=\sigma_1^+\sigma_0^+\sigma_{-1}^+|\Downarrow\rangle$ in Fig. \ref{current}. For particular choice of site $n$, one can identify certain bound states from the currents. In Fig. \ref{current}(a), where the currents are calculated at site $n=15$, the first bump of the currents comes at the same time for different $\Delta$. It is due to the wavefronts of the free magnon state whose propagating speed is independent of $\Delta$. While the second bump corresponding to the two-magnon bound state comes in the order of increasing $\Delta$, since the speed of bound states decreases with $\Delta$. For $\Delta=4$, time is not long enough to see the 
two-magnon bound state arrive. In Fig. \ref{current}(b) we show the current at site $n=9$, closer to the origin. The free state has already passed for all the $\Delta$'s, while the two-magnon bound states can be seen clearly coming in order of increasing $\Delta$. For $\Delta=0.9$, the three-magnon bound state has too small a weight to show in the current, while for $\Delta=1.2$ one can see the three-magnon bound state arriving. We expect that current measurement could be a easier way to probe the bound states. 

{\it Conclusions} We have studied various aspects of the quench dynamics of the XXZ Heisenberg model. We presented in (\ref{wavefunction}) the exact time-dependent wavefunction in a integral form, and showed how it generated the bound state structure of thermodynamic Bethe Ansatz. We evaluated the integral numerically and by  saddle point approximation, and calculated the evolution of observables including local magnetization, staggered magnetization and spin currents. 

{\bf Acknowledgments}
 We are grateful to  E. Altman, J. Lebowitz and in particular to  D. Iyer  for stimulating discussions. This work was supported by NSF grant DMR-1006684 and DMR-1104501.

\bibliography{reference.bib}

\begin{thebibliography}{27}
\expandafter\ifx\csname natexlab\endcsname\relax\def\natexlab#1{#1}\fi
\expandafter\ifx\csname bibnamefont\endcsname\relax
  \def\bibnamefont#1{#1}\fi
\expandafter\ifx\csname bibfnamefont\endcsname\relax
  \def\bibfnamefont#1{#1}\fi
\expandafter\ifx\csname citenamefont\endcsname\relax
  \def\citenamefont#1{#1}\fi
\expandafter\ifx\csname url\endcsname\relax
  \def\url#1{\texttt{#1}}\fi
\expandafter\ifx\csname urlprefix\endcsname\relax\def\urlprefix{URL }\fi
\providecommand{\bibinfo}[2]{#2}
\providecommand{\eprint}[2][]{\url{#2}}

\bibitem[{\citenamefont{Bloch et~al.}(2008)\citenamefont{Bloch, Dalibard, and
  Zwerger}}]{RevModPhys.80.885}
\bibinfo{author}{\bibfnamefont{I.}~\bibnamefont{Bloch}},
  \bibinfo{author}{\bibfnamefont{J.}~\bibnamefont{Dalibard}}, \bibnamefont{and}
  \bibinfo{author}{\bibfnamefont{W.}~\bibnamefont{Zwerger}},
  \bibinfo{journal}{Rev. Mod. Phys.} \textbf{\bibinfo{volume}{80}},
  \bibinfo{pages}{885} (\bibinfo{year}{2008}).

\bibitem[{\citenamefont{Kinoshita et~al.}(2006)\citenamefont{Kinoshita, Wenger,
  and Weiss}}]{kinoshita2006quantum}
\bibinfo{author}{\bibfnamefont{T.}~\bibnamefont{Kinoshita}},
  \bibinfo{author}{\bibfnamefont{T.}~\bibnamefont{Wenger}}, \bibnamefont{and}
  \bibinfo{author}{\bibfnamefont{D.~S.} \bibnamefont{Weiss}},
  \bibinfo{journal}{Nature} \textbf{\bibinfo{volume}{440}},
  \bibinfo{pages}{900} (\bibinfo{year}{2006}).

\bibitem[{\citenamefont{Rigol et~al.}(2007)\citenamefont{Rigol, Dunjko,
  Yurovsky, and Olshanii}}]{PhysRevLett.98.050405}
\bibinfo{author}{\bibfnamefont{M.}~\bibnamefont{Rigol}},
  \bibinfo{author}{\bibfnamefont{V.}~\bibnamefont{Dunjko}},
  \bibinfo{author}{\bibfnamefont{V.}~\bibnamefont{Yurovsky}}, \bibnamefont{and}
  \bibinfo{author}{\bibfnamefont{M.}~\bibnamefont{Olshanii}},
  \bibinfo{journal}{Phys. Rev. Lett.} \textbf{\bibinfo{volume}{98}},
  \bibinfo{pages}{050405} (\bibinfo{year}{2007}).

\bibitem[{\citenamefont{Rigol et~al.}(2008)\citenamefont{Rigol, Dunjko, and
  Olshanii}}]{rigol2008thermalization}
\bibinfo{author}{\bibfnamefont{M.}~\bibnamefont{Rigol}},
  \bibinfo{author}{\bibfnamefont{V.}~\bibnamefont{Dunjko}}, \bibnamefont{and}
  \bibinfo{author}{\bibfnamefont{M.}~\bibnamefont{Olshanii}},
  \bibinfo{journal}{Nature} \textbf{\bibinfo{volume}{452}},
  \bibinfo{pages}{854} (\bibinfo{year}{2008}).

\bibitem[{\citenamefont{Cazalilla}(2006)}]{PhysRevLett.97.156403}
\bibinfo{author}{\bibfnamefont{M.~A.} \bibnamefont{Cazalilla}},
  \bibinfo{journal}{Phys. Rev. Lett.} \textbf{\bibinfo{volume}{97}},
  \bibinfo{pages}{156403} (\bibinfo{year}{2006}).

\bibitem[{\citenamefont{Manmana et~al.}(2007)\citenamefont{Manmana, Wessel,
  Noack, and Muramatsu}}]{PhysRevLett.98.210405}
\bibinfo{author}{\bibfnamefont{S.~R.} \bibnamefont{Manmana}},
  \bibinfo{author}{\bibfnamefont{S.}~\bibnamefont{Wessel}},
  \bibinfo{author}{\bibfnamefont{R.~M.} \bibnamefont{Noack}}, \bibnamefont{and}
  \bibinfo{author}{\bibfnamefont{A.}~\bibnamefont{Muramatsu}},
  \bibinfo{journal}{Phys. Rev. Lett.} \textbf{\bibinfo{volume}{98}},
  \bibinfo{pages}{210405} (\bibinfo{year}{2007}).

\bibitem[{\citenamefont{Barthel and
  Schollw\"ock}(2008)}]{PhysRevLett.100.100601}
\bibinfo{author}{\bibfnamefont{T.}~\bibnamefont{Barthel}} \bibnamefont{and}
  \bibinfo{author}{\bibfnamefont{U.}~\bibnamefont{Schollw\"ock}},
  \bibinfo{journal}{Phys. Rev. Lett.} \textbf{\bibinfo{volume}{100}},
  \bibinfo{pages}{100601} (\bibinfo{year}{2008}).

\bibitem[{\citenamefont{Fioretto and Mussardo}(2010)}]{fioretto2010quantum}
\bibinfo{author}{\bibfnamefont{D.}~\bibnamefont{Fioretto}} \bibnamefont{and}
  \bibinfo{author}{\bibfnamefont{G.}~\bibnamefont{Mussardo}},
  \bibinfo{journal}{New Journal of Physics} \textbf{\bibinfo{volume}{12}},
  \bibinfo{pages}{055015} (\bibinfo{year}{2010}).

\bibitem[{\citenamefont{Kollath et~al.}(2007)\citenamefont{Kollath, L\"auchli,
  and Altman}}]{PhysRevLett.98.180601}
\bibinfo{author}{\bibfnamefont{C.}~\bibnamefont{Kollath}},
  \bibinfo{author}{\bibfnamefont{A.~M.} \bibnamefont{L\"auchli}},
  \bibnamefont{and} \bibinfo{author}{\bibfnamefont{E.}~\bibnamefont{Altman}},
  \bibinfo{journal}{Phys. Rev. Lett.} \textbf{\bibinfo{volume}{98}},
  \bibinfo{pages}{180601} (\bibinfo{year}{2007}).

\bibitem[{\citenamefont{Biroli et~al.}(2010)\citenamefont{Biroli, Kollath, and
  L\"auchli}}]{PhysRevLett.105.250401}
\bibinfo{author}{\bibfnamefont{G.}~\bibnamefont{Biroli}},
  \bibinfo{author}{\bibfnamefont{C.}~\bibnamefont{Kollath}}, \bibnamefont{and}
  \bibinfo{author}{\bibfnamefont{A.~M.} \bibnamefont{L\"auchli}},
  \bibinfo{journal}{Phys. Rev. Lett.} \textbf{\bibinfo{volume}{105}},
  \bibinfo{pages}{250401} (\bibinfo{year}{2010}).

\bibitem[{\citenamefont{Calabrese et~al.}(2011)\citenamefont{Calabrese, Essler,
  and Fagotti}}]{PhysRevLett.106.227203}
\bibinfo{author}{\bibfnamefont{P.}~\bibnamefont{Calabrese}},
  \bibinfo{author}{\bibfnamefont{F.~H.~L.} \bibnamefont{Essler}},
  \bibnamefont{and} \bibinfo{author}{\bibfnamefont{M.}~\bibnamefont{Fagotti}},
  \bibinfo{journal}{Phys. Rev. Lett.} \textbf{\bibinfo{volume}{106}},
  \bibinfo{pages}{227203} (\bibinfo{year}{2011}).

\bibitem[{\citenamefont{Calabrese and Cardy}(2007)}]{1742-5468-2007-06-P06008}
\bibinfo{author}{\bibfnamefont{P.}~\bibnamefont{Calabrese}} \bibnamefont{and}
  \bibinfo{author}{\bibfnamefont{J.}~\bibnamefont{Cardy}},
  \bibinfo{journal}{Journal of Statistical Mechanics: Theory and Experiment}
  \textbf{\bibinfo{volume}{2007}}, \bibinfo{pages}{P06008}
  (\bibinfo{year}{2007}).

\bibitem[{\citenamefont{Rossini et~al.}(2010)\citenamefont{Rossini, Suzuki,
  Mussardo, Santoro, and Silva}}]{PhysRevB.82.144302}
\bibinfo{author}{\bibfnamefont{D.}~\bibnamefont{Rossini}},
  \bibinfo{author}{\bibfnamefont{S.}~\bibnamefont{Suzuki}},
  \bibinfo{author}{\bibfnamefont{G.}~\bibnamefont{Mussardo}},
  \bibinfo{author}{\bibfnamefont{G.~E.} \bibnamefont{Santoro}},
  \bibnamefont{and} \bibinfo{author}{\bibfnamefont{A.}~\bibnamefont{Silva}},
  \bibinfo{journal}{Phys. Rev. B} \textbf{\bibinfo{volume}{82}},
  \bibinfo{pages}{144302} (\bibinfo{year}{2010}).

\bibitem[{\citenamefont{Gobert et~al.}(2005)\citenamefont{Gobert, Kollath,
  Schollw\"ock, and Sch\"utz}}]{PhysRevE.71.036102}
\bibinfo{author}{\bibfnamefont{D.}~\bibnamefont{Gobert}},
  \bibinfo{author}{\bibfnamefont{C.}~\bibnamefont{Kollath}},
  \bibinfo{author}{\bibfnamefont{U.}~\bibnamefont{Schollw\"ock}},
  \bibnamefont{and} \bibinfo{author}{\bibfnamefont{G.}~\bibnamefont{Sch\"utz}},
  \bibinfo{journal}{Phys. Rev. E} \textbf{\bibinfo{volume}{71}},
  \bibinfo{pages}{036102} (\bibinfo{year}{2005}).

\bibitem[{\citenamefont{Pozsgay}(2013)}]{1742-5468-2013-10-P10028}
\bibinfo{author}{\bibfnamefont{B.}~\bibnamefont{Pozsgay}},
  \bibinfo{journal}{Journal of Statistical Mechanics: Theory and Experiment}
  \textbf{\bibinfo{volume}{2013}}, \bibinfo{pages}{P10028}
  (\bibinfo{year}{2013}).

\bibitem[{\citenamefont{Fagotti and Essler}(2013)}]{1742-5468-2013-07-P07012}
\bibinfo{author}{\bibfnamefont{M.}~\bibnamefont{Fagotti}} \bibnamefont{and}
  \bibinfo{author}{\bibfnamefont{F.~H.~L.} \bibnamefont{Essler}},
  \bibinfo{journal}{Journal of Statistical Mechanics: Theory and Experiment}
  \textbf{\bibinfo{volume}{2013}}, \bibinfo{pages}{P07012}
  (\bibinfo{year}{2013}).

\bibitem[{\citenamefont{Barmettler et~al.}(2010)\citenamefont{Barmettler, Punk,
  Gritsev, Demler, and Altman}}]{barmettler2010quantum}
\bibinfo{author}{\bibfnamefont{P.}~\bibnamefont{Barmettler}},
  \bibinfo{author}{\bibfnamefont{M.}~\bibnamefont{Punk}},
  \bibinfo{author}{\bibfnamefont{V.}~\bibnamefont{Gritsev}},
  \bibinfo{author}{\bibfnamefont{E.}~\bibnamefont{Demler}}, \bibnamefont{and}
  \bibinfo{author}{\bibfnamefont{E.}~\bibnamefont{Altman}},
  \bibinfo{journal}{New Journal of Physics} \textbf{\bibinfo{volume}{12}},
  \bibinfo{pages}{055017} (\bibinfo{year}{2010}).

\bibitem[{\citenamefont{Fagotti}(2013)}]{fagotti2013dynamical}
\bibinfo{author}{\bibfnamefont{M.}~\bibnamefont{Fagotti}},
  \bibinfo{journal}{arXiv preprint arXiv:1308.0277}  (\bibinfo{year}{2013}).

\bibitem[{\citenamefont{Bethe}(1931)}]{Bethe1931}
\bibinfo{author}{\bibfnamefont{H.~A.} \bibnamefont{Bethe}},
  \bibinfo{journal}{Z. Phys.} \textbf{\bibinfo{volume}{71}},
  \bibinfo{pages}{205} (\bibinfo{year}{1931}).

\bibitem[{\citenamefont{Ganahl et~al.}(2012)\citenamefont{Ganahl, Rabel,
  Essler, and Evertz}}]{PhysRevLett.108.077206}
\bibinfo{author}{\bibfnamefont{M.}~\bibnamefont{Ganahl}},
  \bibinfo{author}{\bibfnamefont{E.}~\bibnamefont{Rabel}},
  \bibinfo{author}{\bibfnamefont{F.~H.~L.} \bibnamefont{Essler}},
  \bibnamefont{and} \bibinfo{author}{\bibfnamefont{H.~G.}
  \bibnamefont{Evertz}}, \bibinfo{journal}{Phys. Rev. Lett.}
  \textbf{\bibinfo{volume}{108}}, \bibinfo{pages}{077206}
  (\bibinfo{year}{2012}).

\bibitem[{\citenamefont{{Fukuhara} et~al.}(2013)\citenamefont{{Fukuhara},
  {Schau{\ss}}, {Endres}, {Hild}, {Cheneau}, {Bloch}, and
  {Gross}}}]{2013arXiv1305.6598F}
\bibinfo{author}{\bibfnamefont{T.}~\bibnamefont{{Fukuhara}}},
  \bibinfo{author}{\bibfnamefont{P.}~\bibnamefont{{Schau{\ss}}}},
  \bibinfo{author}{\bibfnamefont{M.}~\bibnamefont{{Endres}}},
  \bibinfo{author}{\bibfnamefont{S.}~\bibnamefont{{Hild}}},
  \bibinfo{author}{\bibfnamefont{M.}~\bibnamefont{{Cheneau}}},
  \bibinfo{author}{\bibfnamefont{I.}~\bibnamefont{{Bloch}}}, \bibnamefont{and}
  \bibinfo{author}{\bibfnamefont{C.}~\bibnamefont{{Gross}}},
  \bibinfo{journal}{Nature} \textbf{\bibinfo{volume}{502}}, \bibinfo{pages}{76}
  (\bibinfo{year}{2013}).

\bibitem[{\citenamefont{Yudson}(1988)}]{yudson1988dynamics}
\bibinfo{author}{\bibfnamefont{V.}~\bibnamefont{Yudson}},
  \bibinfo{journal}{Physics Letters A} \textbf{\bibinfo{volume}{129}},
  \bibinfo{pages}{17} (\bibinfo{year}{1988}).

\bibitem[{\citenamefont{Iyer and Andrei}(2012)}]{PhysRevLett.109.115304}
\bibinfo{author}{\bibfnamefont{D.}~\bibnamefont{Iyer}} \bibnamefont{and}
  \bibinfo{author}{\bibfnamefont{N.}~\bibnamefont{Andrei}},
  \bibinfo{journal}{Phys. Rev. Lett.} \textbf{\bibinfo{volume}{109}},
  \bibinfo{pages}{115304} (\bibinfo{year}{2012}).

\bibitem[{\citenamefont{Iyer et~al.}(2013)\citenamefont{Iyer, Guan, and
  Andrei}}]{PhysRevA.87.053628}
\bibinfo{author}{\bibfnamefont{D.}~\bibnamefont{Iyer}},
  \bibinfo{author}{\bibfnamefont{H.}~\bibnamefont{Guan}}, \bibnamefont{and}
  \bibinfo{author}{\bibfnamefont{N.}~\bibnamefont{Andrei}},
  \bibinfo{journal}{Phys. Rev. A} \textbf{\bibinfo{volume}{87}},
  \bibinfo{pages}{053628} (\bibinfo{year}{2013}).

\bibitem[{\citenamefont{Orbach}(1958)}]{PhysRev.112.309}
\bibinfo{author}{\bibfnamefont{R.}~\bibnamefont{Orbach}},
  \bibinfo{journal}{Phys. Rev.} \textbf{\bibinfo{volume}{112}},
  \bibinfo{pages}{309} (\bibinfo{year}{1958}).

\bibitem[{\citenamefont{Sutherland}(2004)}]{sutherland2004beautiful}
\bibinfo{author}{\bibfnamefont{B.}~\bibnamefont{Sutherland}},
  \emph{\bibinfo{title}{Beautiful models: 70 years of exactly solved quantum
  many-body problems}} (\bibinfo{publisher}{World Scientific Publishing
  Company}, \bibinfo{year}{2004}).

\bibitem[{\citenamefont{Takahashi}(2005)}]{takahashi2005thermodynamics}
\bibinfo{author}{\bibfnamefont{M.}~\bibnamefont{Takahashi}},
  \emph{\bibinfo{title}{Thermodynamics of one-dimensional solvable models}}
  (\bibinfo{publisher}{Cambridge University Press}, \bibinfo{year}{2005}).

\end{thebibliography}

\end{document}